\documentclass[prb,superscriptaddress,twocolumn]{revtex4}
\usepackage{graphicx}

\begin{document}
\preprint{Version 8; 2005 June 09}

\title{The origin of switching noise in GaAs/AlGaAs lateral gated devices}

\author{M.~Pioro-Ladri\`{e}re}
\affiliation{Institute for Microstructural Sciences, National Research Council of Canada, Ottawa, Ontario, Canada K1A~0R6}
\affiliation{Centre de Recherche sur les Propri\'{e}t\'{e}s \'{E}lectroniques de Mat\'{e}riaux Avanc\'{e}s, Universit\'{e} de Sherbrooke, Sherbrooke, Qu\'{e}bec, Canada J1K~2R1}

\author{John~H.~Davies}
\email{jdavies@elec.gla.ac.uk}
\affiliation{Department of Electronics and Electrical Engineering, 
Glasgow University, Glasgow, G12~8QQ, U. K.}

\author{A.~R.~Long}
\affiliation{Department of Physics and Astronomy, Glasgow University, Glasgow, G12~8QQ, U. K.}

\author{A.~S.~Sachrajda}
\email{Andy.Sachrajda@nrc-cnrc.gc.ca}
\affiliation{Institute for Microstructural Sciences, National Research Council of Canada, Ottawa, Ontario, Canada K1A~0R6}

\author{Louis~Gaudreau}
\affiliation{Centre de Recherche sur les Propri\'{e}t\'{e}s \'{E}lectroniques de Mat\'{e}riaux Avanc\'{e}s, Universit\'{e} de Sherbrooke, Sherbrooke, Qu\'{e}bec, Canada J1K~2R1}

\author{P.~Zawadzki}
\author{J.~Lapointe}
\author{J.~Gupta}
\author{Z.~Wasilewski}
\author{S.~Studenikin}
\affiliation{Institute for Microstructural Sciences, National Research Council of Canada, Ottawa, Ontario, Canada K1A~0R6}

\date{\today}

\begin{abstract}
We have studied switching (telegraph) noise at low temperature in GaAs/AlGaAs heterostructures with lateral gates and introduced a model for its origin, which explains why noise can be suppressed by cooling samples with a positive bias on the gates. 
The noise was measured by monitoring the conductance fluctuations around $e^2/h$ on the first step of a quantum point contact at around 1.2\,K. 
Cooling with a positive bias on the gates dramatically reduces this noise, while an asymmetric bias exacerbates it. 
Our model is that the noise originates from a leakage current of electrons that tunnel through the Schottky barrier under the gate into the conduction band and become trapped near the active region of the device. 
The key to reducing noise is to keep the barrier opaque under experimental conditions. 
Cooling with a positive bias on the gates reduces the density of ionized donors. 
This builds in an effective negative gate voltage so that a smaller negative bias is needed to reach the desired operating point. 
This suppresses tunnelling from the gate and hence the noise. 
The reduction in the density of ionized donors also strengthens the barrier to tunneling at a given applied voltage. 
Further support for the model comes from our direct observation of the leakage current into a closed quantum dot, around $10^{-20}\,\mathrm{A}$ for this device. 
The current was detected by a neighboring quantum point contact, which showed monotonic steps in time associated with the tunneling of single electrons into the dot. 
If asymmetric gate voltages are applied, our model suggests that the noise will increase as a consequence of the more negative gate voltage applied to one of the gates to maintain the same device conductance. 
We observe exactly this behaviour in our experiments. 
\end{abstract}



\maketitle

\section{Introduction}

Lateral gated devices can be defined in the two dimensional electron gas (2DEG) of a semiconductor heterostructure using surface gates. \cite{cwjb91} 
Their tunability makes them the ideal choice for many fundamental applications, such as the implementation of solid state quantum bits; 
a single electron can be isolated and probed in single and double dot devices. \cite{mc02,mpl03,jme04,jrp04} 
Unfortunately, they are extremely sensitive to fluctuations in their local electrostatic environment. 
These fluctuations provide one of the most important obstacles to developing such devices for future applications in areas such as spin and charge qubits, where a quiet electrostatic environment is essential. 
While fluctuations can result from external sources such as voltage noise on a gate, which is relatively easy to remove, the main difficulties stem from noise which behaves as if it is intrinsic to the wafer. 
Changes in impurity configurations and in the charge states of electronic traps are examples of such noise. 
This type of temporal fluctuation results in a random switching of device characteristics \cite{sm54} known as random telegraph noise (RTN). 
Its consequences include deterioration in the stability of the number of electrons in quantum dots and the conductance of mesoscopic conductors.

Unlike in MOSFET devices, where the source of RTN is better understood, \cite{mjk89,mx04} no definitive conclusions have been reached to explain the origin of switching noise observed in GaAs/AlGaAs 2DEG devices. 
There exists a long history of the study of this noise. \cite{ypl90,gt90,dhc91a,cd91,dhc92a,ck97} 
Several mechanisms have been proposed to account for RTN, as well as the related $1/f$ and Lorentzian noise, including a gate leakage current through localized states, electron trapping, switching events in a remote doping layer and DX centres. 
Nevertheless, it is well established experimentally that the noise level in GaAs/AlGaAs gated devices can be affected by a gate voltage. 
For example $1/f$ noise in submicron Hall devices can be reduced by applying a moderate voltage on the gate \cite{yl04a} and RTN in quantum dot devices can be reduced by applying a positive voltage to the gates during cooldown. \cite{cm05} 
The latter technique, known as bias cooling, provides a ``frozen in'' gate voltage at low temperature related to the filling of DX centres at higher temperatures. \cite{pmm91a,arl93a} 
For structures with a uniform gate, the procedure was shown to affect the degree of correlations established in the doping layer between the positively charged donors and the negatively charged DX centers thereby influencing the 2DEG mobility. \cite{eb94a,ptc97a}

In this paper we present a detailed study of the effect of bias cooling on the noise characteristics of GaAs/AlGaAs gated nanodevices and develop a model to explain these results. 
By carefully monitoring the noise level in quantum point contacts (QPCs) for different bias voltages applied during cool down, we find that the frequency of switching events in moderately noisy samples can be reduced to zero beyond a positive threshold bias. 
In one particularly noisy sample, the noise was reduced by at least two orders of magnitude over a narrow window of positive bias. 
We also observed a pronounced dependence of the noise on the difference in voltage between the gates. 
We propose that the RTN is triggered by a small leakage current of electrons tunneling from the gate into the conduction band of the heterostructure. 
Some electrons are trapped in long-lived localized states near the active region of the device and cause RTN before they reach the 2DEG. 
We explain how bias cooling can suppress this leakage current through its influence on the barrier under the gate. 
Direct evidence for the leakage current, which is an essential feature of our model, is provided by monitoring the charge trapped in a closed quantum dot.

\section{Switching noise in quantum point contacts}
\label{qpc section}

\subsection{Experimental technique}

The heterostructure used in this study was grown by molecular-beam epitaxy and included a 40\,nm thick $\mathrm{Al_{0.33}Ga_{0.67}As}$ spacer, 2 monolayers of GaAs (5.6\,\r{A}), a 40\,nm thick $\mathrm{Al_{0.33}Ga_{0.67}As}$ layer uniformly doped with Si, capped by 15\,nm of GaAs. The doping density was $2.07 \times 10^{18}\,\mathrm{cm}^{-3}$. The 2DEG was thus located at an interface 95.56\,nm below the surface. The density and mobility of the 2DEG were $1.7 \times 10^{11}\,\mathrm{cm}^{-2}$ and $2 \times 10^6 \,\mathrm{cm}^2 \,\mathrm{V}^{-1} \,\mathrm{s}^{-1}$ respectively. 

\begin{figure}
\includegraphics{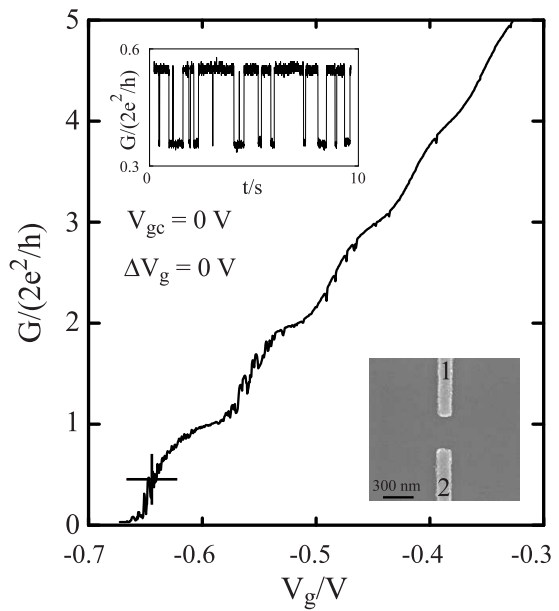}
\caption{Conductance $G$ of the quantum point contact measured in sample 1 as a function of gate voltage $V_\mathrm{g}$. $V_\mathrm{g}$ is defined as the average gate voltage applied to the gates, $V_\mathrm{g} = {1 \over 2} (V_\mathrm{g1} + V_\mathrm{g2})$. The trace was taken with symmetric gate voltages so that the difference $\Delta V_\mathrm{g} = V_\mathrm{g1} - V_\mathrm{g2} = 0$ after cooling the sample with zero bias voltage ($V_\mathrm{gc} = 0$) applied on the gates. 
Top inset: Conductance as a function of time $t$ with $V_\mathrm{g}$ set to the value marked by the cross. 
Lower inset: SEM picture of the device showing the two gates used to define the QPC in the 2DEG.}
\label{qpc cond fig}
\end{figure}

To characterize RTN in gated structures, we fabricated quantum point contact devices. The QPC was employed as a simple but sensitive tool for monitoring changes in the local electrostatic potential caused by charge fluctuations. Using the conductance of the QPC's rather than Coulomb blockade peaks in lateral dots as a noise monitor has the advantage that dilution refrigerator temperatures are not required. A SEM picture of the two gates used to form the QPC is shown as an inset to Fig.~\ref{qpc cond fig}. By applying negative voltage on the gates a one-dimensional channel is formed in the 2DEG whose conductance $G$ can be tuned by the gate voltages; $V_\mathrm{g1}$ and $V_\mathrm{g2}$ are the voltages to the gates labeled 1 and 2. The conductance as function of gate voltage was measured using a 4-terminal lock-in technique while the noise was monitored in time by using 2-terminal dc current measurements. All measurements were performed in a 1.2\,K pumped helium cryostat.

The bias cooling procedure involved the application of a common bias, $V_\mathrm{gc}$, to all gates while cooling the samples from room temperature to 1.2\,K over a 30 minute period. In order to reinitialize the device, the sample was illuminated for one minute with a light-emitting diode at room temperature between each thermal cycle.

\subsection{Results}
\label{results section}

All gated samples fabricated on the wafer showed RTN. Some devices were, however, much noisier than others. We assume, following previous work, \cite{ypl90,gt90,dhc91a,cd91,dhc92a,ck97} that the conductance of the QPC is modulated by the electrostatic potential from electrons trapped nearby. The QPC conductance reflects changes in the occupation of these traps. The time dependence and magnitude of the current flowing through the QPC provides a fingerprint of these fluctuations. 

Fig.~\ref{qpc cond fig} shows the evolution of $G$ for a typical sample as the average $V_\mathrm{g}$ of the two gate voltages is swept. The difference between the two gate voltages $\Delta V_\mathrm{g}$ is zero. For this particular cooldown and sample, only one trap was active. The sweep rate was chosen to be slower than the mean switching frequency of the fluctuations. At a temperature of 1.2\,K, 4 steps of $2e^2/h$ in the conductance are clearly resolved. Each time an electron is trapped (de-trapped) by the defect, a step down (up) in conductance is observed. The amplitude of the jumps depends on the sensitivity (given by the slope $dG/dV_\mathrm{g}$) and is the highest between plateaus. A close inspection of the curve reveals that the frequency of jumps increases as the gate voltages are made more negative, confirming that the occupation of the trap is affected by the gate voltage. For small gate voltages, the trap occupation (QPC conductance) does not fluctuate. In this regime the QPC has the higher conductance, enabling us to conclude that the trap is empty at equilibrium (for zero applied gate voltage). 

To characterize the noise, the gate voltages were set to maximize the sensitivity of the QPC. A dc bias of $500\,\mu \mathrm{V}$ was applied between the source and drain contacts and the amplified drain current was monitored using a oscilloscope. The inset in Fig.~\ref{qpc cond fig} shows a time trace of the two-terminal dc conductance taken at the value of $V_\mathrm{g}$ marked by the cross. The random telegraphic switches between empty and occupied states of the trap are clearly observed. By taking many similar traces, characteristics including the average time the trap remains empty or occupied and the mean frequency of level changes could be extracted. This way of characterizing noise is simple and did not require sophisticated spectrum analysis. It is well suited to devices with a relatively low number of traps, where individual changes in states can be resolved in time.

\begin{figure}
\includegraphics{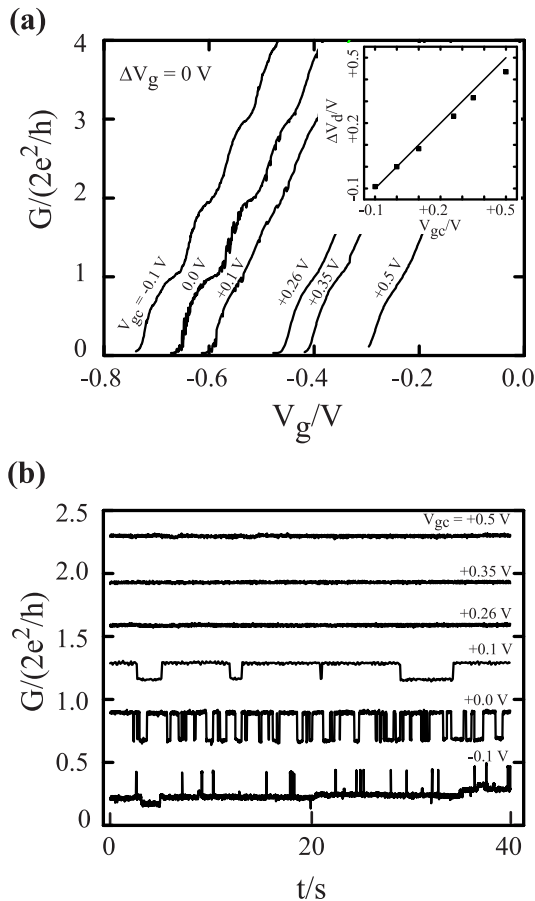}
\caption{Effect of bias cooling on sample 1. 
(a)~QPC gate voltage characteristic for several bias voltages $V_\mathrm{gc}$ applied during cooldown. 
Inset: Shift $\Delta V_\mathrm{d}$ of the depletion threshold measured for each cooling bias. The line is a guide to the eye showing a shift equal to the applied bias. 
(b)~Time traces taken at maximum sensitivity for each bias. The traces are shifted vertically for clarity.}
\label{bias cool cond fig}
\end{figure}

We now turn to the effect of bias cooling. Fig.~\ref{bias cool cond fig}(a) shows the QPC gate voltage characteristic for different cooldown bias voltages $V_\mathrm{gc}$. For a positive (negative) bias, the gate voltage characteristic shifts towards less (more) negative voltages. In the inset, the shift of the depletion point is plotted as a function of the cooling bias, i.e.\ the gate voltage value at which the channel is formed, with respect to the zero bias cooldown value. The line corresponds to a shift that equals the applied bias. One can see that for small biases, a built-in voltage forms during cooldown which is very close to $-V_\mathrm{gc}$. \cite{arl93a} For higher bias there is a very small deviation. A similar behavior is observed for the other threshold features such as the QPC pinch off. 

Bias cooling has a large effect on the switching noise: The QPC becomes dramatically quieter as $V_\mathrm{gc}$ becomes more positive. This is more clearly seen in Fig.~\ref{bias cool cond fig}(b) where time traces for the correspondent bias cooling values of Fig.~\ref{bias cool cond fig}(a) are plotted. Each curve was taken with the QPC set to maximum sensitivity ($G \approx e^2/h$). For this particular sample, the effect of a positive bias cool was to reduce the frequency of level changes to zero monotonically. For $V_\mathrm{gc} \ge +0.26\,\mathrm{V}$, the conductance shows no signs of telegraph noise during the entire time span of measurements. Conversely, the application of a negative bias seems to activate additional traps. 

\begin{figure}
\includegraphics{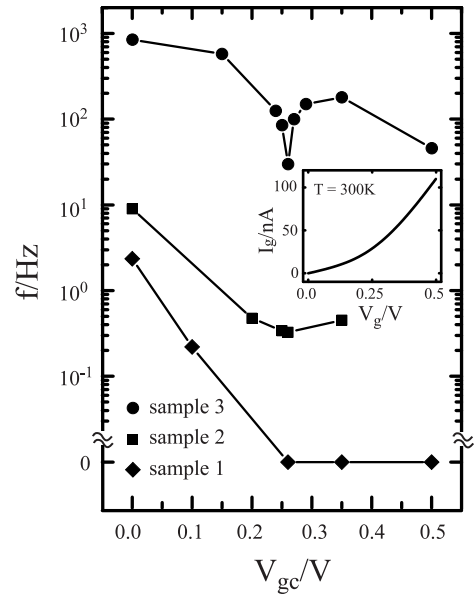}
\caption{Mean frequency $f$ of the switching noise as a function of cooling bias $V_\mathrm{gc}$ for 3 different samples fabricated on the same wafer. The lowest curve was obtained on sample 1, and has $f = 0$ for $V_\mathrm{gc} > 0.26\,\mathrm{V}$. Inset: Leakage current $I_\mathrm{g}$ as a function of $V_\mathrm{g}$ measured at room temperature for sample 2.}
\label{noise frequency fig}
\end{figure}

Fig.~\ref{noise frequency fig} shows the mean frequency of level changes extracted from the time traces as a function of cooldown bias for three samples. For the two quieter samples, bias cooling reduced the noise until $V_\mathrm{gc} \approx 0.25\,\mathrm{V}$; at larger bias sample 1 showed no switching at all while the noise in sample 2 remained roughly constant at a low value. 
The noisiest sample (3) showed a notch centered at $+0.26\,\mathrm{V}$, where the level of noise is reduced by almost two orders of magnitude. 

The leakage characteristics of the gates at high temperature are shown in the inset. At low temperature this current is too small to be measured on a picoammeter but can be detected indirectly, as we shall show in Sec.~\ref{dot section}. 

\begin{figure}
\includegraphics{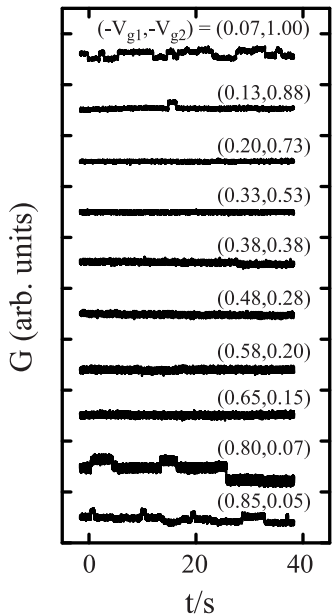}
\caption{Effect of gate voltage asymmetry. Traces of $G$ as a function of time taken for different positive and negative value of the gate voltage difference $\Delta V_\mathrm{g}$ after cooling sample 1 with a positive bias $V_\mathrm{gc} = +0.26\,\mathrm{V}$ on both gates. Traces are offset for clarity. For each trace, the voltages (in $\mathrm{V}$) applied on gates 1 and 2 used to set the QPC to maximum sensitivity are indicated in parentheses.}
\label{delta vg fig}
\end{figure}

The noise characteristics depend not only the chosen cooling bias but also on the voltage difference $\Delta V_\mathrm{g} = V_\mathrm{g1} - V_\mathrm{g2}$ between the two QPC gates. Fig.~\ref{delta vg fig} shows time traces taken for several positive and negative $\Delta V_\mathrm{g}$ values after cooling sample 1 with $+0.26\,\mathrm{V}$. Each trace was obtained by setting the QPC to its maximum sensitivity. For $\Delta V_\mathrm{g} = 0\,\mathrm{V}$ ($V_\mathrm{g1} = V_\mathrm{g2} = -0.38\,\mathrm{V}$), the sample is quiet as expected. As $\Delta V_\mathrm{g}$ is increased, the sample remains quiet for absolute values below around 0.7\,V. Above this threshold, the RTN switches on for both positive and negative $\Delta V_\mathrm{g}$.

\section{Model to explain switching noise and the effect of bias cooling}
\label{model section}

A mechanism that explains the observed effect of bias cooling on the noise must satisfy these two, apparently conflicting, constraints.
\begin{enumerate}
\item
The primary effect of a gate bias $+V_\mathrm{gc}$ during the cool is to alter the ionization of donors near the surface; it has no direct effect on the 2DEG. After the bias is removed at low temperature the sample behaves as if there were a built-in gate bias of $V_\mathrm{g} = -V_\mathrm{gc}$. 
\item
The fluctuations are caused by a charge that must be trapped close enough to the QPC to modulate the conductance by around $0.4 e^2/h$, and with a lifetime of around 1\,s. Thus the trap is probably in the spacer layer or channel. 
\end{enumerate}
A change in ionization near the \emph{surface} must therefore affect the occupation of trapped electrons near the \emph{channel}. We propose that there is a small leakage current of electrons that tunnel from the gate to the channel. Some of these electrons become trapped close to the QPC, causing the fluctuations. \cite{dhc91a} This current is limited by the Schottky barrier under the gate, which is strongly affected by the bias cool through its influence on the local electrostatic field. We shall now explain this in detail.

\subsection{Effect of bias cooling}

\begin{figure}
\includegraphics[scale=0.36]{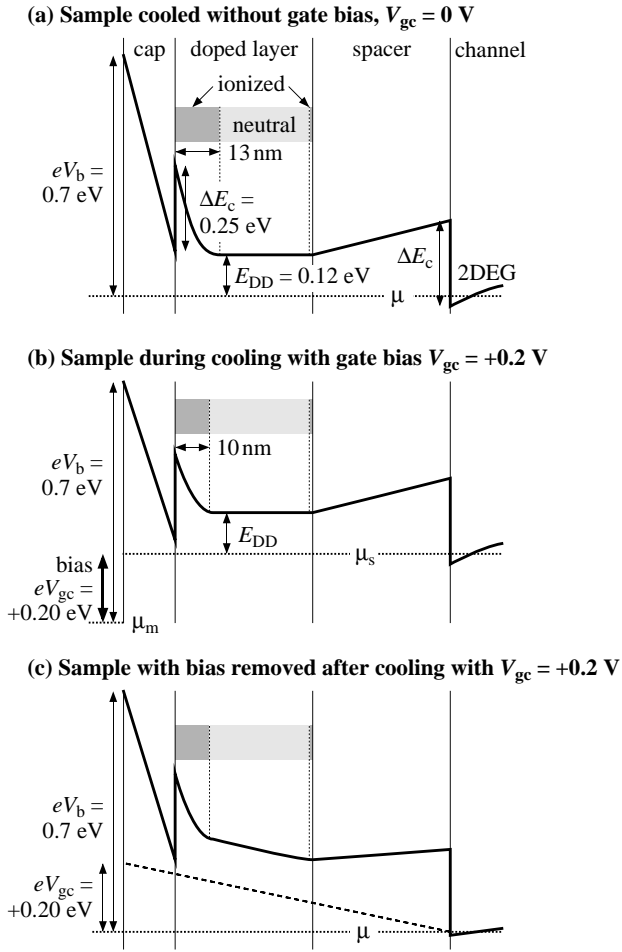}
\caption{Profiles of conduction band under a large gate 
(a)~for a sample cooled without gate bias, 
(b)~during cooling with a gate bias of $V_\mathrm{gc} = +0.2\,\mathrm{V}$, and 
(c)~after removing the bias applied in (b). 
The chemical potential of the gate $\mu_\mathrm{m}$ is pinned at an energy $eV_\mathrm{b}$ below the GaAs conductance band by the surface states, while the chemical potential of the semiconductor $\mu_\mathrm{s}$ is pinned by DX centers at an energy $E_\mathrm{DD}$ below the conduction band in the neutral region of the doped layer.}
\label{bias cool bands fig}
\end{figure}

Fig.~\ref{bias cool bands fig}(a) shows the conduction band as a function of depth under a large gate on a sample that has been cooled at equilibrium ($V_\mathrm{gc} = 0$). \cite{arl93a,jhd98z} An important feature is that the doped layer has separated into three regions.
\begin{enumerate}
\item
A thick, shallow, ionized layer next to the cap generates the potential needed for the Schottky barrier on the surface, which holds the conduction band at an an energy $eV_\mathrm{b}$ above the Fermi level $\mu_\mathrm{m}$ of the gate.
\item
The middle of the doped layer is neutral. The electrons are all trapped in DX levels, \cite{pmm91a} which pin the conduction band at an energy $E_\mathrm{DD}$ above the Fermi level $\mu_\mathrm{s}$ of the semiconductor. This sample has $\mu_\mathrm{m} = \mu_\mathrm{s} = \mu$ because no bias is applied.
\item
There is a thin, deep, ionized layer next to the spacer to balance the charge of the 2DEG. It is less than 1\,nm thick.
\end{enumerate}
For simplicity the sketches are drawn on the assumption of an abrupt transition between ionized and neutral regions. The neutral region is present because the wafer contains far more donors than are needed to generate the electrostatic potentials at the surface and 2DEG. This is typical and is vital to our model. We concentrate here on layers where the doping is spread over a slab but the model also works for $\delta$-doping. \cite{es95a}

Now consider the sample shown in Fig.~\ref{bias cool bands fig}(b) during cooling with a bias $V_\mathrm{gc} = +0.2\,\mathrm{V}$ on the gate. Donors are free to change their occupation at room temperature when the bias is first applied. The mobile charge closest to the gate responds to the bias, which therefore attracts electrons into the doped layer. This reduces the thickness of the shallow layer of ionized donors next to the cap from 13\,nm to 10\,nm. There is no effect on the 2DEG at all. 

The occupation of the donors becomes fixed when the temperature of the sample is lowered through 100\,K because the barriers to trapping and de-trapping become too high. \cite{arl93a} Thus the distribution of donors shown in Figs.~\ref{bias cool bands fig}(a) and (b) is ``frozen'' into the sample at low temperature. 
Fig.~\ref{bias cool bands fig}(c) shows the effect of removing the bias of $V_\mathrm{gc} = +0.2\,\mathrm{V}$ after the material has been fully cooled: it is as if a bias of $V_\mathrm{g} = -0.2\,\mathrm{V}$ has been superimposed on the sample in Fig.~\ref{bias cool bands fig}(b). 
The only mobile charge is now the 2DEG in the channel, whose density is reduced by the apparent bias. 
The argument that a bias cool at $V_\mathrm{gc}$ becomes equivalent to a built-in gate bias of $V_\mathrm{g} = -V_\mathrm{gc}$ at low temperature applies equally to patterned gates. 
The shift of the threshold voltage for the QPC was plotted in the inset to Fig.~\ref{bias cool cond fig}(a), where the data lie very close to the predicted line of unit slope.

\subsection{Leakage current}

\begin{figure}
\includegraphics[scale=0.36]{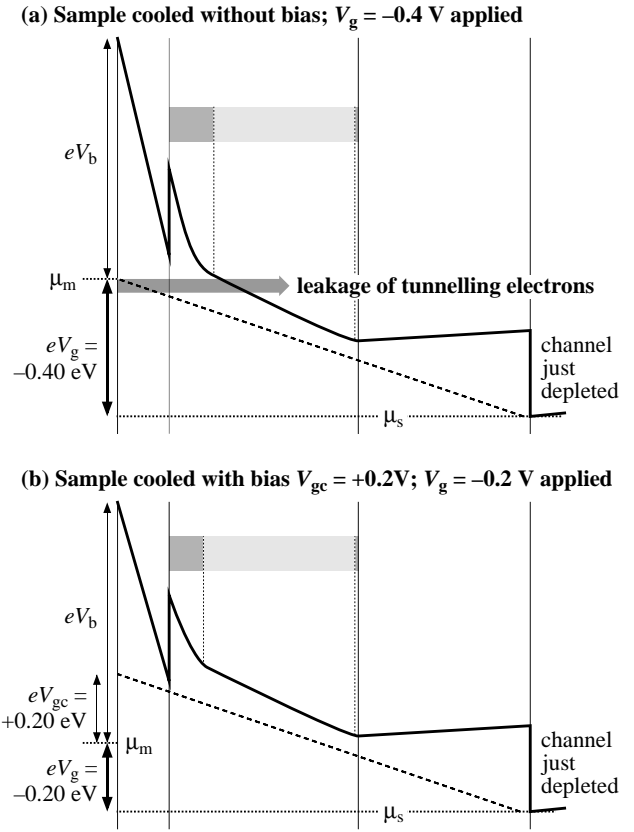}
\caption{(a)~Conduction band profile for a gate voltage of $V_\mathrm{g} = -0.4\,\mathrm{V}$ applied to a sample cooled without bias. Electrons are able to tunnel from the gate into the doped layer. 
(b)~Conduction band profile for the same operating point after a bias cool of $V_\mathrm{gc} = +0.2\,\mathrm{V}$. 
Because of the built-in gate voltage, only $V_\mathrm{g} = -0.2\,\mathrm{V}$ is needed to achieve the same effective gate voltage of $-0.4\,\mathrm{V}$.
Tunneling into the doped layer is no longer possible. }
\label{cutoff bands fig}
\end{figure}

Now consider a negative bias applied to patterned gates so that a QPC such as the one shown in the inset to Fig.~\ref{qpc cond fig} is just cut off. 
Suppose that the design requires $V_\mathrm{g} = -0.4 \,\mathrm{V}$ for cutoff on layers that were cooled without bias; this small magnitude was chosen to make the sketches clear. 
Fig.~\ref{cutoff bands fig}(a) shows the conduction band along a line that goes from a gate to the channel at the midpoint of the QPC (this line is not normal to the surface as in Fig.~\ref{bias cool bands fig}). 
The QPC is only just cut off so the conduction band in the midpoint of the channel brushes the Fermi level $\mu_\mathrm{s}$. 
The negative bias on the gate raises its Fermi level $\mu_\mathrm{m}$ by 0.4\,eV, which permits electrons to tunnel into the channel. 
This gives rise to a leakage current and ultimately to the switching noise. 

The leakage can be classified into three regimes.
\begin{enumerate}
\item
The current is tiny for small negative bias because electrons must tunnel through the full thickness of the cap, doped and spacer layers into the channel.
\item
The current starts to increase rapidly when when the bias rises to $-eV_\mathrm{g} > \Delta E_\mathrm{c} \approx 0.25\,\mathrm{eV}$, because electrons need no longer tunnel through the spacer. 
\item
As the bias rises further, the barrier for electrons at the Fermi level $\mu_\mathrm{m}$ becomes narrower until electrons need tunnel only through the cap and the shallow ionized layer of donors, less than 30\,nm.
The sample in Fig.~\ref{cutoff bands fig}(a) is just in this limit.
This happens roughly when $-eV_\mathrm{g} > 3 E_\mathrm{DD} \approx 0.35\,\mathrm{eV}$ due to the ``lever-arm factor''.
Any further increase in bias has a smaller effect.
\end{enumerate}
The rapid growth of the leakage current with increasing negative bias explains the observation in Sec.~\ref{results section} that the noise tends to increase as more negative bias is applied.

In contrast, Fig.~\ref{cutoff bands fig}(b) shows the comparable situation for layers that were cooled under a bias of $V_\mathrm{gc} = +0.2\,\mathrm{V}$. 
There is a built-in bias of $-0.2\,\mathrm{V}$ from the cooling, so an applied bias of only $V_\mathrm{g} = -0.2\,\mathrm{V}$ is needed to reach the same operating point as in Fig.~\ref{cutoff bands fig}(a). 
The Fermi level in the gate $\mu_\mathrm{m}$ is therefore raised by only 0.2\,eV instead of 0.4\,eV. 
Leakage remains in regime 1 with tunneling all the way to the channel, nearly 100\,nm. 

Thus a positive bias cool suppresses tunneling because a smaller applied bias is needed for the same operating point. 
Moreover, bias cooling also reduces tunneling at the \emph{same} applied bias. 
This is because the smaller density of ionized donors causes the conduction band to fall more slowly away from the Schottky barrier $eV_\mathrm{b}$ under the gate. 
Figs.~\ref{bias cool bands fig}(a) and (c) show this at equilibrium: 
the barrier is more opaque in the bias cooled sample (c) and this feature is preserved when a negative bias is applied to the gate.

We have calculated the effect of bias cooling on the current density $J$ from a large gate biased to the threshold voltage, using a simple WKB estimate of tunneling through uniform layers as in Fig.~\ref{bias cool bands fig}. 
\begin{itemize}
\item
For $V_\mathrm{gc} = 0$ the conditions at threshold lie on the crossover between regimes 1 and 2 above, giving $J \approx 10^{-16} \,\mathrm{A}\,\mathrm{m}^{-2}$. 
\item
Positive bias cooling with $V_\mathrm{gc} = +0.2\,\mathrm{V}$ brings the sample firmly into regime 1, where electrons must tunnel from the gate all the way to the channel, and $J$ falls to $10^{-57} \,\mathrm{A}\,\mathrm{m}^{-2}$. 
\item
Negative bias cooling with $V_\mathrm{gc} = -0.1\,\mathrm{V}$ leads to a narrow barrier, just the cap and shallow ionized layer of donors (regime 3). Leakage rises to $10^{-6} \,\mathrm{A}\,\mathrm{m}^{-2}$, which is consistent with our observations in Sec.~\ref{dot section}. 
\end{itemize}
Although the real situation in our three-dimensional devices is more complex than for this simple model, our calculation shows that the effect of bias cooling on the tunnel current is dramatic because the geometry of the barrier at threshold is greatly changed. 
A stronger bias is needed to deplete a device with a patterned gate and larger values of $V_\mathrm{gc}$ are therefore needed to suppress tunneling.

\subsection{Origin of switching noise}

Electrons that tunnel from the gate must eventually reach the 2DEG but may take many routes. 
If they travel entirely in the conduction band there will be no switching noise; the observed fluctuations require a trap close to the QPC. 
Cobden \textit{et al.} \cite{dhc92a} observed an irreversible telegraph noise signal and argued that transport was by hopping between localized states. 
Our signals appear to arise from independent traps and usually only one is active. 
We suggest that most of the transport occurs in the conduction band, although the final state before the 2DEG must be a sufficiently deep trap to give fluctuations of the observed frequency.

We feel that it is unlikely that electrons are trapped in their passage through most of the doped layer, despite the high density of possible traps provided by the donors. 
Shallow levels would become ionized in the high electric field under the gate, while electrons that enter the deep DX levels would probably be trapped permanently at the low temperature of the experiment. 
Their charge would accumulate and cause a slow shift of threshold voltages with time. 
Such drifts have not been observed in our experiments.

The localized state that gives rise to the switching noise is most likely to be in the thin layer of ionized donors next to the spacer layer. 
The hydrogenic level of a donor gives a binding energy of around 6\,meV, which is reinforced by the V-shaped potential well generated by the layer of charge, shown in Fig.~\ref{cutoff bands fig}. 
Electrons must tunnel through the 40\,nm spacer to reach the channel and a very rough estimate of the lifetime is not inconsistent with the timescales observed in the experiments. 
These localized states would be present even in a perfectly clean sample. 
However, there may also be defects or impurities in the nominally undoped spacer or channel, which could provide further telegraph signals. 
On the other hand, if their lifetime is short, such impurities might instead reduce noise by providing a ``short circuit'' across the localized states among the donors. 
A resonant channel of this sort could explain the notch seen in the top curve in Fig.~\ref{noise frequency fig}.

In most cases an electron will reach an open region of 2DEG, from which it can return to the contacts. 
However, it is also possible that it might enter a confined region of 2DEG, such as a quantum dot. 
We shall explore this in Sec.~\ref{dot section}.

Our model also explains other features of the experiments. 
The effect of bias cooling saturates for $V_\mathrm{gc} > 0.3\,\mathrm{V}$ because the conduction band in the cap is pulled below the Fermi level in the semiconductor, $\mu_\mathrm{s}$. 
This results in a third channel of electrons in the cap that screens the donors from any further change in bias. 
Secondly, a device with asymmetrically biased gates is noisier than when the same average bias is applied symmetrically, because an increase of the negative bias on one gate increases the transparency of the barrier underneath.

\section{Experimental evidence for leakage current}
\label{dot section}

\begin{figure}
\includegraphics{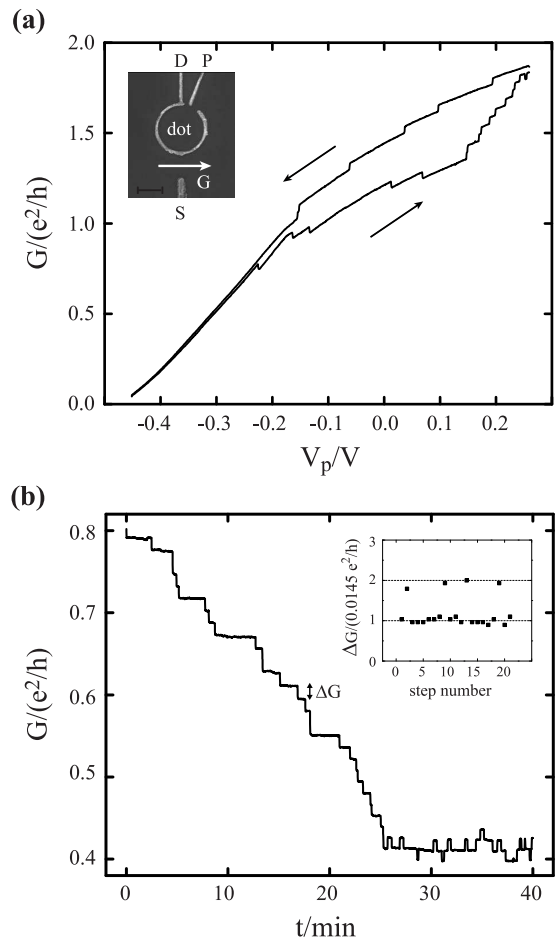}
\caption{(a)~Conductance $G$ of the QPC defined between gates D and S as a function of the voltage $V_\mathrm{P}$ applied to gate P. $V_\mathrm{P}$ is first swept from $+0.26\,\mathrm{V}$ to $-0.45\,\mathrm{V}$ (top curve) and then back to $+0.26\,\mathrm{V}$ (bottom curve). 
Inset: SEM picture of the device showing the labels of the gates. The dot is $0.7\,\mu\mathrm{m}$ in diameter. 
(b)~Evolution of the sensor conductance with time $t$ after ramping $V_\mathrm{P}$ quickly from $+0.26\,\mathrm{V}$ to 0\,V. 
Inset: Normalized step height $\Delta G$ plotted for the first 20 charging events.}
\label{dot cond fig}
\end{figure}

We now provide direct evidence for the leakage current by setting up a situation where electrons tunneling from the gate enter a closed region, which acts as a ``corral''. This is achieved by a quantum dot that can be isolated from the contacts together with a QPC charge sensor. Electrons that enter this region cannot be collected by the contacts and thus the number of electrons in the closed region should increase with time. The inset of Fig.~\ref{dot cond fig}(a) shows a SEM micrograph of the device. Similar observations were made on two devices. By applying a negative voltage on gate D, a quantum dot of circular geometry is formed within the 2DEG. Gate P is used to control the tunneling rate between the dot and the contacted 2DEG. This occurs under the gap in gate D. Gate S is used to form the QPC. Because the QPC channel is close to the dot, it is sensitive to changes in its charge state. \cite{mf93,mak04,jme03,jrp04} The device was bias cooled in a dilution refrigerator with a positive bias of $+0.26\,\mathrm{V}$ applied on all three gates. The QPC and the dot were defined by applying a symmetric bias on gate S and D of $-0.604\,\mathrm{V}$. As described above, this voltage is sufficient for a leakage current to flow between the gates and the 2DEG. 

Fig.~\ref{dot cond fig}(a) shows the QPC conductance as gate P is swept >from $+0.26\,\mathrm{V}$ towards a more negative value and back again. It might be expected that a negative bias would repel electrons from the dot, in which case an upward step in conductance would be seen in the QPC. However, the opposite is seen: The sweep in the negative direction causes discrete \emph{downward} steps of equal height. The smooth variation between steps is caused by the direct electrostatic coupling between the QPC channel and the gate. We attribute each step to a single electron leaking from gate D and finding its way into the quantum dot. When $V_\mathrm{P}$ is ramped back, additional charging events are observed until a critical voltage of $V_\mathrm{P} \approx +0.15\,\mathrm{V}$ is reached. At this point, the tunneling link between the dot and the 2DEG is re-opened and the electrons that were collected in the dot can leave. Each discharging event is picked up as a step up in the QPC conductance. There are 11 steps down and 10 up; one electron remains trapped in the dot, giving a small difference in conductance of the QPC. 

To monitor this effect as a function of time, $V_\mathrm{P}$ was quickly ramped from equilibrium ($+0.26\,\mathrm{V}$) to 0\,V where the tunneling link was pinched off. Fig.~\ref{dot cond fig}(b) shows the resulting evolution of the sensor conductance with time. In the first 25 minutes, several charging events are observed, each corresponding to the addition of one extra electron to the quantum dot. In the inset, we plotted the change in conductance associated to each charging event. Following that, the charge state of the quantum dot fluctuates by plus or minus one electron suggesting that a steady state is reached in which the gate, dot and 2DEG form a ``Coulomb blockade'' type of circuit. From the number of observed events and assuming that for two electrons tunneling from gate D, only one on average ends up in the dot, we estimate a leakage current due to this gate of 10\,zA ($1\,\mathrm{zA} = 10^{-21}\,\mathrm{A}$) in this regime.

\section{Conclusion}

We have shown that the level of switching (telegraph) noise in a AlGaAs/GaAs gated device can be dramatically reduced by applying a positive gate bias while the device is cooled down and have provided a model to explain this. 
Such a cooling bias was also found to shift the characteristics of the device as if there were a built-in gate voltage of equal magnitude but opposite sign, as we predict. 
Our model is that the cooling bias reduces the density of ionized donors by freezing free carriers into DX centres in the doping layer, thus shifting the threshold voltage. 
We propose that the noise arises from a leakage current of electrons that tunnel from the gate into the conduction band. 
Before they reach the channel they are trapped near the active region of the device, probably in a thin layer of ionized donors next to the spacer, where their Coulomb potential modulates the conductance. 
Cooling with a positive bias reduces this leakage in two ways: 
the built-in potential reduces the applied bias needed to achieve a given operating point, and the reduced density of ionized donors enhances the barrier to tunneling. 
We found direct evidence for the existence of the leakage current by monitoring the charge in a closed quantum dot. 
The model described in this study provides an important handle for controlling the noise in lateral quantum devices made with surface gates and for future designs of semiconductor wafers.

\begin{acknowledgments}
We acknowledge useful discussions with D.~G.~Austing, P.~Hawrylak and C.~Marcus. 
A.S.S. is grateful for support from CIAR. 
M. P.-L. acknowledges NSERC and FQRNT for financial support.
This collaboration was stimulated by the Canadian--European Research Initiative on Nanostructures CERION-2. 
\end{acknowledgments}

\end{document}